\documentclass[11pt]{article}
\usepackage{graphicx}
\usepackage{comment}
\usepackage{amsmath,amssymb}
\usepackage{ragged2e}
\usepackage{cite,xspace}
\usepackage{comment,array,soul} 
\usepackage{longtable}
\usepackage{xparse}
\NewExpandableDocumentCommand\mcc{O{1}m}
    {\multicolumn{#1}{l}{#2}}
\usepackage{multirow}

\def\Title#1{\begin{center} {\Large #1 } \end{center}}
\def\Author#1{\begin{center}{ \sc #1} \end{center}}
\def\Address#1{\begin{center}{ \it #1} \end{center}}

\newcommand\pubblock{\begin{flushright}   
SMU-HEP-23-03
\\ \pubdate  \end{flushright}}
\newenvironment{Abstract}{\begin{quotation}  }{\end{quotation}}
\newenvironment{Presented}{\begin{quotation} \begin{center} 
             PRESENTED AT\end{center}\bigskip 
      \begin{center}\begin{large}}{\end{large}\end{center} \end{quotation}}

\usepackage[a4paper,total={170mm,257mm},left=20mm,top=20mm]{geometry}

\newcommand\pubdate{August 27, 2023}

\newcommand{\orcid}[1]{\,\href{https://orcid.org/#1}{\includegraphics[width=9pt]{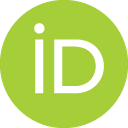}}}

\newcommand{\orcidFO}{0000-0001-6799-2436} 
\newcommand{\orcidPN}{0000-0003-3732-0860} 
\newcommand{\orcidAC}{0000-0001-8906-2440} 
\newcommand{\orcidMAX}{0009-0003-0139-4072} 
\newcommand{\orcidLK}{0009-0007-5639-0350} 

\usepackage{url} 
\usepackage{hyperref}
\usepackage{xcolor} 
\usepackage{multirow}
\definecolor{nicered}{rgb}{0.5,0.,0.}
\definecolor{nicegreen}{rgb}{0.,0.5,0.}
\definecolor{niceblue}{rgb}{0.,0.,0.5}
\definecolor{darkpink}{rgb}{0.8,0.47,0.47}
\hypersetup{colorlinks,citecolor=nicegreen,linkcolor=nicered,urlcolor=niceblue}
\usepackage{enumitem} 
\usepackage[normalem]{ulem}
\setlist{nolistsep} 
\usepackage{setspace}
\usepackage{bibspacing}

\definecolor{nicered}{rgb}{0.5,0.,0.}
\definecolor{nicegreen}{rgb}{0.,0.5,0.}
\definecolor{niceblue}{rgb}{0.,0.,0.5}
\definecolor{darkpink}{rgb}{0.8,0.47,0.47}
\hypersetup{colorlinks,citecolor=nicegreen,linkcolor=nicered,urlcolor=niceblue}
\usepackage{enumitem} 
\usepackage[normalem]{ulem}
\setlist{nolistsep} 
 \usepackage{setspace}

\newcommand{\fanto}{{\sf Fant\^omas}\xspace}
\newcommand{\metam}{metamorph\xspace}
\newcommand{\xFitter}{{\sf xFitter}\xspace}

\begin{document}
\renewcommand{\thefootnote}{\fnsymbol{footnote}}
\begin{titlepage}

\pubblock
\vfill

\Title{\fanto For QCD: parton distributions in a pion with Bézier parametrizations}
\vfill

\Author{Aurore Courtoy\orcid{\orcidAC}\footnote{Email: aurore@fisica.unam.mx} and Maximiliano Ponce-Chavez\orcid{\orcidMAX}}
\Address{Instituto de F\'isica,
  Universidad Nacional Aut\'onoma de M\'exico, Apartado Postal 20-364,
  01000 Ciudad de M\'exico, Mexico\looseness=-1}
\Author{Lucas Kotz\orcid{\orcidLK}\footnote{Presenter; email: lkotz@smu.edu}, Pavel Nadolsky\orcid{\orcidPN}\footnote{Email: nadolsky@smu.edu}, Fred Olness\orcid{\orcidFO}, \& Varada Purohit}
\Address{Department of Physics, Southern Methodist University,
  Dallas, TX 75275-0181, USA\looseness=-1}

\vfill
 \begin{Abstract}
We report on a new framework to parametrize parton distribution functions (PDFs) and other hadronic nonperturbative functions using polynomial functions realized by B\'ezier curves. B\'ezier parameterizations produce a stable fit with a low number of free parameters, while competing in performance with neural networks and offering explicit interpretation. We specifically apply this approach to determine PDFs in a pion, essential for understanding  of nonperturbative QCD dynamics. 
 \end{Abstract}

\vfill
\begin{Presented}
DIS2023: XXX International Workshop on Deep-Inelastic Scattering and
Related Subjects, \\
Michigan State University, USA, 27-31 March 2023 \\
     \includegraphics[width=9cm]{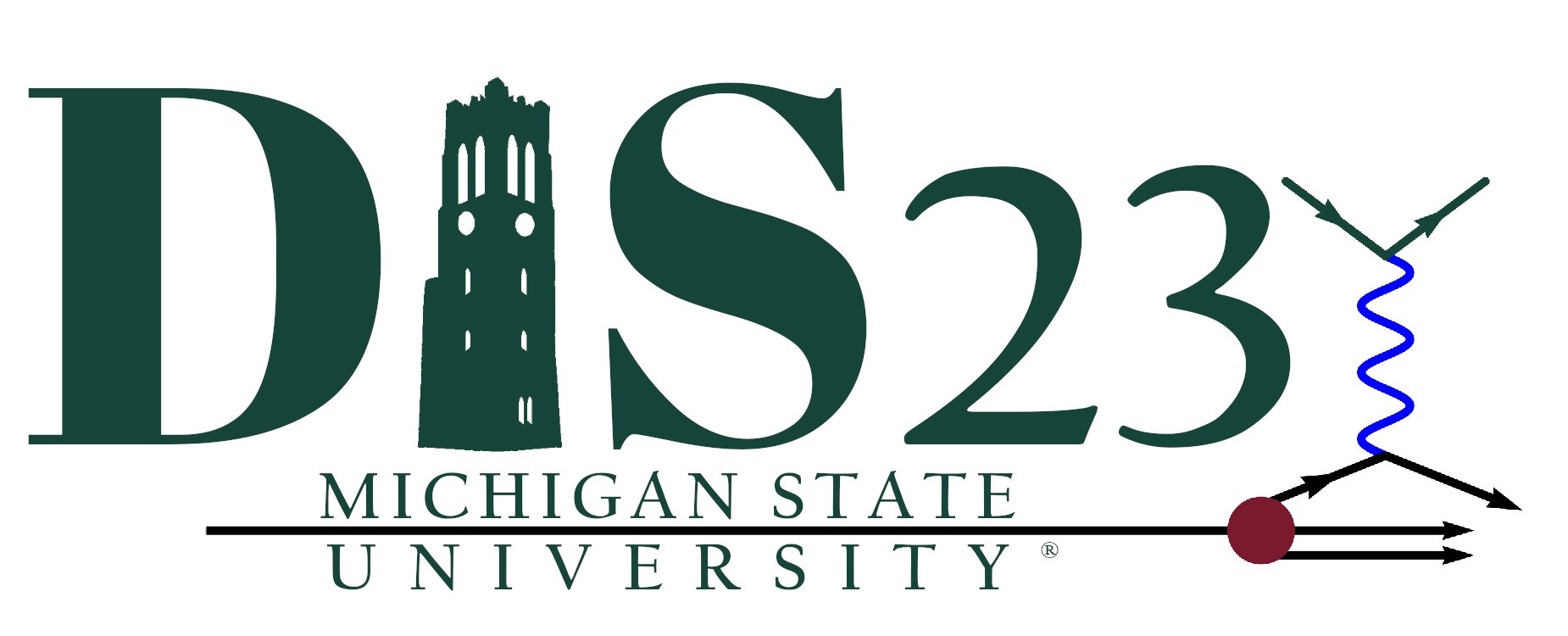}
\end{Presented}
\vfill

\end{titlepage}
\renewcommand{\thefootnote}{\arabic{footnote}}
\setcounter{footnote}{0}

\section{Introduction}

The structure of hadrons has been a subject of active research in quantum chromodynamics (QCD) for several decades. Numerous experiments have been conducted to probe hadrons/nuclei with either a lepton, deep-inelastic scattering (DIS), or other hadrons, {\it e.g.} Drell-Yan process (DY). Hadronic cross sections for these experiments involve convolutions of collinear parton distribution functions (PDFs), commonly extracted from the data using log-likehood minimization or a related method. In many such analyses, PDFs are parameterized at the initial evolution scale $Q_0$ using analytic functional forms, or alternatively they are parameterized using neural networks. Our contribution focuses on parametrization dependence of PDFs in charged pions, which have attracted recent attention \cite{Novikov:2020snp, Barry:2018ort, Barry:2021osv}. In precision QCD analyses, uncertainty due to the choice of the parametric form can be as important as experimental and theoretical uncertainties \cite{Kovarik:2019xvh,Hou:2019efy,Courtoy:2022ocu}. This uncertainty is also central for comparisons with calculations of PDFs using nonperturbative and lattice QCD [see, {\it e.g.},~\cite{RuizArriola:2002bp, Gao:2022iex,Cichy:2021lih,Lu:2022cjx,Rinaldi:2022dyh}]. Our goal is to improve handling of uncertainties when discriminating among theoretical models for pion and other PDFs.

How should one parametrize the PDFs when their functional dependence is only partly known from first principles? While neural networks can approximate any continuous function, in practice polynomial parametrizations with relatively few free parameters are competitive by offering both controlled flexibility and ease of interpretation.  A computational framework to automate construction of PDF parametrizations using polynomials of arbitrary degree thus offers advantages both for finding the best fit of the data and quantifying the uncertainty on the resulting PDFs. 

We developed a C++ program that realizes such framework, named \fanto  For QCD. The functional form of the PDFs in this framework, which we call a {\it \metam}, reproduces the asymptotic limits of PDFs when the momentum fraction $x$ tends to 0 or 1. At intermediate $x$, the functional form can reproduce arbitrary variations by utilizing a polynomial approximation realized using a B\'ezier curve. The \metam function is fully parametrized by the power laws in the asymptotic limits $x\to 0$ and 1, as well as by the function's values at user-chosen {\it control points} in $x$. The \metam parametrization streamlines the choice of the fitted parameters, and it can approximate a wide variety of analytic behaviors as a consequence of ``polynomial mimicry" reviewed in Ref.~\cite{Courtoy:2020fex}.

Given their ability to approximate arbitrary functions with a relatively low yet adjustable number of free parameters, polynomial forms are an attractive option to parametrize the PDFs. Indeed, as discussed in Ref.~\cite{Kovarik:2019xvh}, the number of free parameters must be selected to avoid unstable (too many parameters) or rigid (too few parameters) fits. Repeating a fit with a bundle of polynomial functional forms, while also applying the cross validation techniques, serves the same objective as neural network parametrizations~\cite{NNPDF:2021njg}, namely, exploring a wide range of parametric forms compatible with the fitted data. On the other hand, the explicit, low number of parameters in the polynomial parametrizations offers advantages for epistemic uncertainty quantification compared to ML/AI techniques, as the PDF dependence over a large (preferably, smaller) number of parameters needs to be robustly explored with either technique \cite{Courtoy:2022ocu}. The \metam parameterization aims to address both needs, by mimicking the flexible nature of neural networks while also providing the benefits of polynomial approximation and Hessian techniques for uncertainty quantification. This approach is suitable for approximating many types of nonperturbative functions. 

As a demonstration, we extracted pion PDFs in the \xFitter framework \cite{Alekhin:2014irh} using our \metam parameterization. \xFitter is an open-source program that performs a global analysis of QCD data to determine PDFs. Past fits are already implemented into \xFitter, {\it e.g.} for the pion~\cite{Novikov:2020snp}, allowing for easy access to perform these fits with variation of settings, such as the implementation of the \metam parameterization. This allows us to benchmark our work with the \xFitter's work more easily while ensuring that only the functional form of the PDF differs between the fits.

The structure of pions is particularly instructive:  thanks to the pion's two-body composition and its pseudo-Goldstone boson nature, the pion can be described in effective theories of QCD with low-energy degrees of freedom. Confronting such nonperturbative predictions with observations in pion scattering at higher energies faces conceptual and technical challenges associated with the transition to the perturbative regime. Phenomenological studies of pion scattering observables will hopefully lead to more insights on these issues, which in turn requires robust quantification of the epistemic uncertainty associated with the PDF functional forms.

\begin{figure}
    \centering
    \includegraphics[width=12.cm]{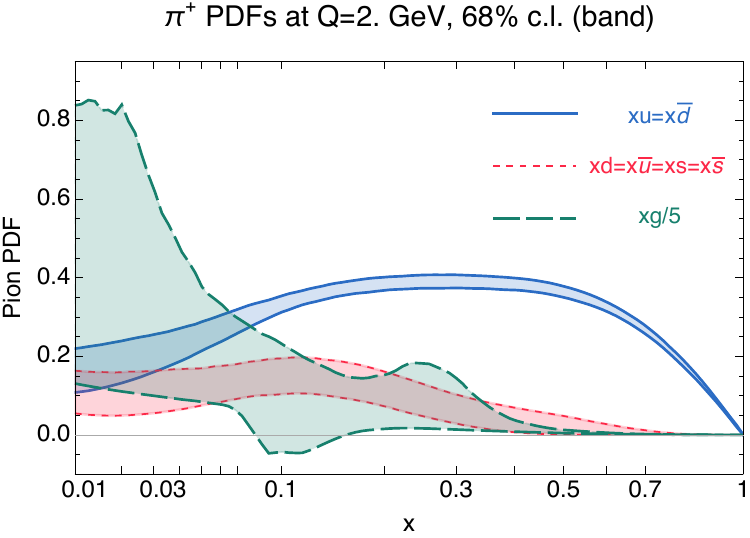}
    \caption{ The $\pi^+$ valence $x u(x, Q^2)=x \bar{d}(x, Q^2)$, sea, and gluon  FantoPDF error bands at $Q=2$~GeV, respectively in blue (solid lines), red (short-dashed lines) and green (long-dashed lines).}
    \label{fig:fin}
\end{figure}

The currently available data for the pion structure are limited compared to those  for the nucleon.  With the metamorph parameterization, we utilize the flexibility of the polynomials to probe the available pion data while taking advantage of the fact that a relatively small number of parameters is sufficient in this case. A variety of settings [see the next sections] were explored, and the final FantoPDF pion Hessian error PDF set is composed using the {\sf METAPDF} method \cite{Gao:2013bia} from the input error PDF sets that differ the most.  The final FantoPDF error bands for the positive pion are plotted, at $Q=2$ GeV, in Fig.~\ref{fig:fin}.

\section{Metamorph parametrizations}

\begin{figure}[ht]
    \centering
    \includegraphics[width=7.8cm]{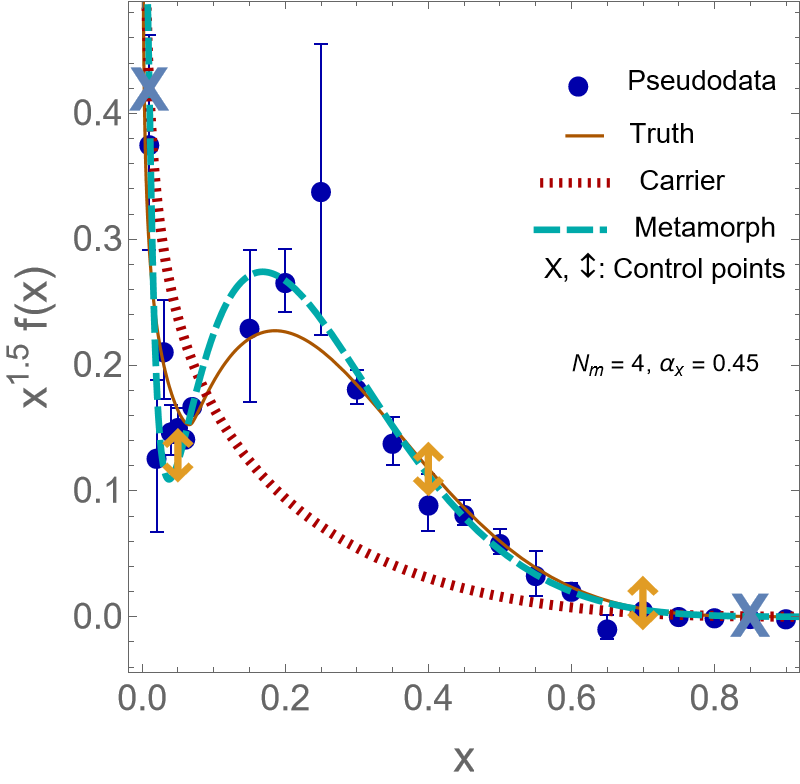}
    \caption{A toy model. The truth function (solid red) is a bimodal function with $x^{-1.5}(1-x)^4\sin(x \pi)$ at low $x$ and $x^{-1.75}(1-x)^4$ at large $x$. The set of pseudodata (blue) is normally distributed around the truth. The carrier (dashed red) specifies the asympotitic behavior for $x\to 0$ and $x\to 1$. The metamorph (dashed cyan) is fitted to the pseudodata using the form in Eq.~(\ref{eq:DISmetafunc}). The metamorph coincides with the carrier asymptotically as a consequence of setting two fixed control points (cyan {\bf X}) at the endpoints of the metamorph. The free control points (orange $\Updownarrow$) are varied to replicate the behavior of the truth in the range $0<x<1$.}
    \label{fig:DIStoyfit}
\end{figure}

Let us now describe the methodology behind the Fant\^omas fits.
To reflect some expected PDF properties, the usual factorization of a functional form is adopted,
\begin{equation}
    f(x)=f_{\mbox{\scriptsize car}}(x)\times f_{\mbox{\scriptsize mod}}(x^{\alpha_x})\,.
    \label{eq:DISmetafunc}
\end{equation}
$f_{\mbox{\scriptsize car}}(x)$ is called the {\it carrier function}, $f_{\mbox{\scriptsize mod}}(x^{\alpha_x})$ is the {\it modulator function}, and $\alpha_x >0$ is a stretching parameter that optimizes placement of control points over $x$. The carrier factor ensures that the PDF obeys expected asymptotic limits as $\lim_{x\to0/1}f(x)$.  These asymptotic limits are determined by the quark-counting rule at $x\to 1$ and by Regge dynamics at $x\to0$. 
The function we chose to use as our carrier is $f_{\mbox{\scriptsize car}}(x)=A_fx^{B_f}(1-x)^{C_f}$.

The modulator function quantifies deviations from the carrier  at $0<x<1$. It reduces to 1 at $x\to 0$ and 1. 
In the \metam parameterization created by the  Fant\^omas For QCD group,
the modulator function values at several control points are parameterized [in lieu of polynomial coefficients directly], allowing the minimizer to replicate a variety of PDF behaviors at $0<x<1$ and find excellent fits with just a few control points.
The modulator function is a polynomial of degree $N_m$ and contains a B\'ezier curve $B_{N_m}(y)$,
\begin{equation}
    \mathcal{B}_{N_m}(y)=\sum_{l=0}^{N_m}C_l B_{N_m,l}(y),
    \label{eq:DISBeziercurve}
\end{equation}
where $B_{N_m,l}(y)$ are Bernstein basis polynomials,
\begin{equation}
    B_{N_m,l} (y)\equiv \left(\begin{matrix} N_m  \\ l  \end{matrix}\right) y^l (1-y)^{N_m-l},
\end{equation}
$C_l$ are the B\'ezier coefficients, and $y=x^{\alpha_x}$. Mathematical properties of the B\'ezier curves relevant for the PDF analysis are reviewed in Ref.~\cite{Courtoy:2020fex}. In particular, we must provide $N_{m}+1$ control points when using a polynomial of degree $N_m$ to have a unique solution that goes through all control points. 
Unlike a traditional polynomial form, the coefficients $C_l$ in Eq.~(\ref{eq:DISBeziercurve}) are derived by solving a matrix equation from a vector of function values at the control points provided as an input~\cite{pomax,Courtoy:2020fex}. 

The Bernstein basis polynomials have some advantages over orthogonal polynomials,  {\it e.g.}, Chebyshev ones. Bernstein polynomials are positive-definite and do not oscillate towards the end points. These two properties allow the B\'ezier curve to be a top-tier contender for the PDF parameterization, since it is less likely to cause unwanted behavior in the PDFs, such as a quickly oscillating function. A crucial detail to notice is that, even while the Bernstein polynomials are positive-definite, there is no such restriction on $C_l$. The B\'ezier coefficients can take on any value, which can result in an overall negative PDF. Positivity can be imposed on the metamorph parametrization as an additional condition. 

The separate components of the \metam are illustrated on the example of a toy fit in Fig. \ref{fig:DIStoyfit}. 
 For a certain ``truth" function, we generated a set of pseudodata that are normally distributed around the truth in accord with preselected $1\sigma$ uncertainties. To these data, we fitted a \metam function with the free parameters defined as the shift from the initial parameters of the input function.

The two control points on each extreme are marked with an ``{\bf X}" to mark that the control point is fixed to the value of the carrier function, $P_{i, \mbox{\scriptsize fixed}}=f_{{\mbox{\scriptsize car}}}(x_i)$. By fixing two of the control points at $x\sim0$ and $x\sim1$, the \metam function satisfies the asymptotic behaviors imposed at these limits of $x^{B_f}$ and $(1-x)^{C_f}$. 
At moderate $x$, the modulator can cause the \metam to significantly deviate from the carrier function. Control points in that region are marked by an up-down arrow ``{\bf $\Updownarrow $}." As a result, the \metam parametrization can get closer to the pseudodata and the truth than the carrier function could have. Besides the control points, the agreement between the \metam and pseudodata can be controlled by both $N_m$ and $\alpha_x$ parameters. 

\section{Analysis of pion PDFs with metamorphs in xFitter}

Our package \fanto realizing such PDF parametrizations has been implemented into {\sf xFitter}. Detailed documents on the functionality of \fanto and on our pion analysis will be released in the future.
 To determine the pion PDFs, we followed the same procedures as the analysis of the charged pion PDF by the \xFitter development team ~\cite{Novikov:2020snp}. We reproduced the published \xFitter pion PDFs using the \metam parametrizations (in this case evaluated with $N_m=0$). We added the pion DIS data from HERA to constrain the gluon PDF at small $x$, as described below, and  we explored a variety of options for the \metam parametrization (available to the user), such as fixing the control point to the carrier, fixing the carrier itself or releasing it. In particular, the choice of the degree of the polynomial and the stretching factor have affected the overall shapes and PDF uncertainties.  

Similar to the \xFitter study, our pool of data involves the pion-nuclear Drell-Yan (DY) pair production from E615, NA10, as well as the prompt photon data WA70 that helps to constrain the gluon. However, this selection results in a large error on the gluon at low $x$, which cannot be disentangled from the quark sea PDFs with DY-like data only. To help remedy the issue of poorly constrained gluon and sea PDFs at low $x$, we included 29 points of the leading-neutron production in DIS (LN) data. We used the H1 prescription for these points \cite{H1:2010udv} and a flux prescription based on the light-cone representation [see, {\it e.g.}, Refs.~\cite{Holtmann:1994rs,Hobbs:2014fya} for more evaluations of the pion flux]. Altogether, our final data selection encompasses three above scattering processes to improve the parton flavor separation.

\begin{figure}
    \centering
    \includegraphics[width=0.5\textwidth]{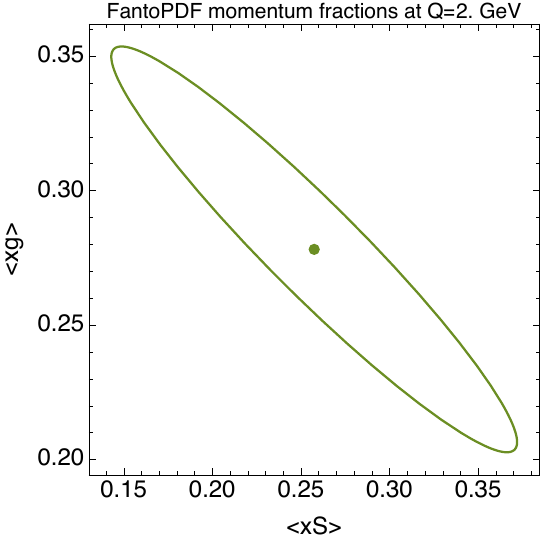}
    \caption{ Correlation ellipse for the momentum fractions of the sea and gluon  FantoPDF  at $Q=2$~GeV. The center of the ellipse corresponds to the central value of the MC interval of the final MC set.}
    \label{fig:correlation}
\end{figure}

 The \metam results can be compared to JAM's results on the pion PDF from 2018 \cite{Barry:2018ort} and 2021 \cite{Barry:2021osv}, as well as to \xFitter's original result~\cite{Novikov:2020snp}. Overall, we find:
 \begin{itemize}
     \item the uncertainties of the FantoPDF set are larger than JAM's or \xFitter's, as they account for sampling over parametrization [a consequence of epistemic uncertainties];
     \item the sea and gluon distribution are indeed determined better at small-$x$ values once LN data are added, but the extrapolation region around $x\sim 0.1$ admits negative solutions for the gluon;
     \item we do not observe a distinct shift (compared to the quoted uncertainties) in the gluon and sea momentum fractions with the addition of LN data, in contrast to the JAM findings~\cite{Barry:2018ort}; instead, the main effect of the LN data is to reduce the PDF uncertainties on the momentum fractions; Fig.~\ref{fig:correlation} shows the anti-correlation of the sea and gluon momentum fractions for the \fanto PDFs; the final result for the \fanto PDF (DY+p$\gamma$+LN) at $Q=2\mbox{ GeV}$ are, evaluated for symmetric $68\%$ C.L. of the final MC set,
     $\langle xf_v \rangle= 0.48\pm 0.05$,  $\langle xf_s \rangle=0.26\pm 0.11$, $\langle xf_g \rangle=0.28\pm 0.08$ 
     for the valence, sea and gluon, respectively, to be compared to the \xFitter results without scale variation [$\langle xf_v \rangle=0.5 \pm 0.02$,  $\langle xf_s \rangle=0.25 \pm 0.08$ and  $\langle xf_g \rangle=0.25 \pm 0.08$] and the range of JAM analyses [for which the gluon momentum fraction has grown to reach $\gtrsim 40\%$ for newer fits]; 
    \item The $\chi^2/N_{\rm pts}$ values for the included fits range between  $1.08\ (440/408)$ and $1.10\ (451/408)$. 
  \end{itemize}

\section{Conclusion}

The internal structure of the pion in high-energy scattering remains relatively unknown, given the paucity of relevant experimental measurements. This will change when the Electron-Ion Collider (EIC) and other future experiments come online. The new data will confront predictions of nonperturbative and lattice QCD methods at energies of order $\sim 1$ GeV with phenomenological PDFs determined from the data using the global QCD analysis. Discrimination of theoretical models in such comparisons requires robust uncertainty quantification, including the epistemic uncertainty associated with the functional forms of PDFs. We have proposed an advanced polynomial parametrization for PDFs based on the B\'ezier curves. In this approach, the bias due to the choice of the parametrization is circumvented by automatically generating multiple functional forms. We nevertheless include some known prior expectations about the PDFs, notably about the asymptotic behaviors at $x\to 0$ or $1$. The structure of the metamorph is amenable to including other prior constraints, e.g. imposing the positivity of PDFs. Even with the similar data as in the recent JAM and \xFitter analyses, the resulting FantoPDF parametrizations predict wider uncertainty bands than in the original publications utilizing a single choice of functional forms. Further details will be reported in upcoming publications.

\section*{Acknowledgments}

The \fanto program developers are grateful for the financial support from 
the Inter-American Network of Networks of QCD Challenges,
a National Science Foundation AccelNet project. This research was also  supported by CONACyT-- Ciencia de Frontera 2019 No.~51244 (FORDECYT-PRONACES). AC and MPC are further supported by the UNAM Grant No. DGAPA-PAPIIT IN111222.
Research at SMU was partially supported by the U.S. Department of Energy under Grant No.~DE-SC0010129. 

\bibliographystyle{utphys}
\bibliography{biblio4fantomas}

\end{document}